\documentstyle[amssymb,prl,aps,epsf]{revtex}

\large

\begin{document}
\title{Universal aspects of vacancy-mediated disordering dynamics: the effect of
external fields.}
\author{Wannapong Triampo$^{1}$, Timo Aspelmeier$^{2}$, and Beate Schmittmann$^{1}$}
\address{$^{1}$Physics Department and Center for Stochastic Processes in Science and
Engineering, \\
Virginia Polytechnic Institute and State University, Blacksburg, VA\\
24061-0435, USA\\
$^{2}$Institut f\"{u}r Theoretische Physik,\\
Universit\"{a}t G\"{o}ttingen,\\
Bunsenstrasse 9, D-37073 G\"{o}ttingen, Germany.\\
}
\date{\today}
\maketitle

\begin{abstract}
We investigate the disordering of an initially phase-segregated binary
alloy, due to a highly mobile defect which couples to an electric or
gravitational field. Using both mean-field and Monte Carlo methods, we show
that the late stages of this process exhibit dynamic scaling, characterized
by a set of exponents and scaling functions. A new scaling variable emerges,
associated with the field. While the scaling functions carry information
about the field and the boundary conditions, the exponents are universal.
They can be computed analytically, in excellent agreement with simulation
results.
\end{abstract}


\vspace{5mm}

\section{Introduction}

The kinetics of phase ordering in alloys, following a rapid temperature
quench below the coexistence curve, has traditionally attracted much
interest, both in the physics and material science communities \cite{reviews}%
. Remarkably, the temporal growth of ordered domains, as measured, e.g., via
a time-dependent structure factor, exhibits universal features, such as
characteristic growth exponents and dynamic scaling, in analogy to critical
phenomena. The ``inverse'' problem, i.e. the bulk {\em disordering} of a
system after a rapid {\em increase} in temperature or in response to highly
mobile defects, is also an interesting problem, occurring in the context of
erosion or corrosion phenomena \cite{er+corr}. In addition, it is a testing
ground for a fundamental question of statistical physics, namely, how a
system approaches its final steady state, starting from an initial {\em %
non-stationary} condition.

In most real solids, microscopic atom-atom exchanges are mediated by
defects, such as vacancies or interstitial sites \cite{basic MS}. Thus, it
is natural to describe such processes with the help of a three-state model,
allowing for two species of atoms and a certain number of empty (defect)
sites \cite{3statemodel}. Vacancy concentrations are typically extremely
small (of the order of $10^{-5}$). As a first step towards a deeper
understanding of disordering dynamics, we will neglect any asymmetries in
the microscopics of the two atom species, leaving them for consideration at
a later stage \cite{STZ}. Thus, our study will be based on an Ising model
with a very small admixture of empty sites, namely a {\em single} vacancy,
undergoing an ``{\em upquench}'' from {\em zero} to {\em infinite} temperature.
Since we wish to describe particles and holes, the dynamics will conserve
the number of each species separately. Only particle-hole exchanges will be
permitted, in order to model the vacancy mechanism. We should note for
completeness that a number of investigations have focused on {\em %
downquenches} in similar models, i.e., vacancy-mediated domain growth \cite
{VMO}.

An alternate view of our study addresses the effect of a random walker on
its background medium. Each step of the walker displaces part of the
background, leaving tracks like a beach comber in the sand. These tracks can
be monitored and display their own dynamics. In the simplest case \cite{TKSZ}%
, the walker is a Brownian vacancy, exploring a lattice filled with
particles of two species, labelled as ``black'' and ``white'', or ``up'' and
``down'' spins. Starting from a perfectly phase-segregated state, three
distinct temporal regimes are observed, separated by two crossover times.
During the late stages of this process, the number of broken bonds exhibits
dynamic scaling, characterized by a set of exponents and a scaling function.
A mean-field theory allows for the analytic calculation of these features,
in excellent agreement with simulations \cite{TKSZ}.

In order to test the range of universality of this model, we extend these
studies to the case of a {\em biased} random walker: The vacancy or defect
hops preferentially along one of the lattice directions, as if subject to
gravity. In the language of electrostatics, the defect is charged,
disordering an initial configuration of neutral atoms in the presence of an
electric field. We emphasize that this is {\em completely equivalent }to
having a neutral (or less massive) vacancy in a background of particles, all
of which carry the {\em same} charge (or larger mass). As in the unbiased
case, there is no {\em feedback} from the background to the defect, i.e.,
the motion of the vacancy is independent of the local background
configuration. However, the choice of boundary conditions along the
direction of the field now becomes important: we will consider two cases,
one in which spatial inhomogeneities persist in the long-time limit, and
another which approaches a homogeneous steady state. Intriguingly, in both
cases the three temporal regimes noted before are still clearly observable.
Our key goal is to explore to what extent scaling functions and exponents
are {\em universal}, i.e., independent of boundary conditions and bias.
Employing exact results, a mean-field theory and Monte Carlo simulations, we
find that the scaling {\em exponents} are completely {\em universal}, but
that the scaling {\em functions} can depend sensitively on the boundary
conditions, through a new scaling variable which involves the bias.

The paper is organized as follows. In Section II, we define our model,
discuss the associated boundary conditions and introduce control parameters
and observables of interest. We then turn to our findings. In Section III,
we set the scene by characterizing the final steady states which are exactly
known. On the basis of Monte Carlo data, we then demonstrate the emergence
of three temporal regimes and the separation of time scales between them.
This observation forms the basis of a mean-field theory, to be introduced in
Section V. Its scaling predictions are compared to detailed Monte Carlo
simulations in the following section. We conclude with some comments and
open questions.

\section{The Model\label{sec2}.}

We begin by summarizing the discrete model underlying the Monte Carlo
simulations. It is defined on a two-dimensional (2D) square lattice of
dimensions $L\times L$. Each lattice site is denoted by a pair of integers, $%
{\bf r\equiv }(x,y)$, and can be occupied by either a black particle, a
white particle or a vacancy. Multiple occupancy is forbidden. Thus the
configurations of the model can be described by a set of spin variables $%
\{\sigma _{{\bf r}}\}$ which can take three values: ${\sigma _{{\bf r}}=+1}$
($-1$) for a black (white) particle, and ${\sigma _{{\bf r}}=0}$ for a
vacancy. To model the {\em minute} concentrations of vacancies in real systems, we
focus on the case of a {\em single} hole. The number of
black $(N^{+})$ and white $(N^{-})$ particles is conserved and equal: $%
N^{+}=N^{-}=\frac{N}{2}$ \cite{o+e}. In the conclusions, we return briefly
to the question of higher vacancy concentrations.

The initial configuration is perfectly phase-segregated: Black and white
particles each fill one half of the system, with a sharp flat interface
between them, chosen to lie horizontally along the $x$-axis. The vacancy is
located at the interface. Even though this structure is that of a
ferromagnetically ordered state, the details of the particle-particle
interactions are not important, since we will be interested in upquenches to
infinite temperature, $T=\infty $. Thus, the vacancy moves independently of
its local environment. However, it is subject to a bias $E$, aligned {\em %
transverse} to the initial interface, i.e., upwards along the (positive) $y$%
-axis. This is of course equivalent to the particles experiencing an
external gravitational or electric field pointing along $-y$. We stress that
a factor $1/T$ has been absorbed into the parameter $E$; i.e., we are
considering the {\em high-temperature, high-field limit} of a more complex
interacting system.

At each Monte Carlo time step (MCS), one of the four nearest neighbors of
the vacancy is chosen at random. The exchange is performed using Metropolis 
\cite{Met} rates: $\min \left\{ 1,e^{Ea\delta y}\right\} $, modelling a {\em %
local} electrostatic or gravitational potential. Here, $\delta y=0,\pm 1$ is
the change in the $y$-coordinate of the vacancy, in units of the lattice
constant $a$. Thus, moves {\em against} the field are exponentially
suppressed while all others are accepted. No particle-particle exchanges are
allowed.

Next, we specify the boundary conditions. In combination with the bias, they
determine whether any spatial inhomogeneities survive in the long-time
limit, with potential consequences for scaling exponents or functions. In
the unbiased case, the stationary state is completely uniform \cite{TKSZ},
and the choice of boundary conditions affects at most nonuniversal
amplitudes. To test the effect of spatial inhomogeneities, we consider two
scenarios, namely, reflecting (also referred to as ``brick wall'', BWBC) or
periodic (PBC) boundary conditions for the top and bottom edges. The right
and left boundaries, being aligned with the drive, are not expected to play
a significant role; we choose them to be periodic in both cases. These two
scenarios differ in two important respects. First, reflecting boundary
conditions allow some spatial inhomogeneities to persist, while periodic
boundary conditions lead to homogeneous distributions. Second, we will see
in the next section that, under BWBC, the system approaches an{\em \
equilibrium} state in which all transport currents have subsided. In
contrast, periodic boundary conditions are incompatible with a global
Hamiltonian, so that the steady state is a {\em non-equilibrium} one,
carrying a net mass current.

The two control parameters for the Monte Carlo simulations will be the field
strength, $E$, and the system size, $L$. To monitor the evolution of the
system, we measure a ``disorder parameter'', namely, the number of broken
bonds (i.e., black-white nearest-neighbor pairs), ${\cal A}(L,E;t)$ \cite
{TKSZ}, as a function of time $t$. More detailed information is carried by
the local hole and magnetization densities, defined respectively as 
\begin{eqnarray}
\phi ({\bf r},t) &\equiv &\left\langle \delta _{{\sigma _{{\bf r}},}%
0}\right\rangle \text{ }  \nonumber \\
\psi ({\bf r},t) &\equiv &\left\langle \sigma _{{\bf r}}\right\rangle \text{ 
}  \label{dens}
\end{eqnarray}
Here, $\left\langle \cdot \right\rangle $ denotes the configurational
average. The Kronecker-$\delta $ ensures that lattice site ${\bf r}$ is the
one occupied by the vacancy. Non-zero values of $\psi ({\bf r},t)$ indicate
an {\em excess} of white or black particles at lattice site ${\bf r}$, which
is also a sensitive measure of the disordering process. The full time
dependence of these densities can in general only be computed within a
mean-field approach. However, their stationary forms are easily found from
the known steady-state distributions, as we will presently see.

We finally note that, for the analytic part of our work, it is
straightforward to generalize our model to $d$ dimensions: Denoting a
lattice site by ${\bf r\equiv }(x_{1},...,x_{d-1},x_{d}\equiv y)$, the field
selects a {\em one-dimensional} subspace, with reflecting or periodic
boundary conditions, along the $y$-direction. The $(d-1)$-dimensional
transverse subspace retains periodic boundaries. 

\section{Exact results: the steady states.}

An exact solution of our model involves knowledge of the full time-dependent
distributions, $P(\{\sigma _{{\bf r}}\},t)$, for the probability of finding
configuration $\{\sigma _{{\bf r}}\}$ at time $t$. This requires finding 
{\em all} eigenvalues and eigenvectors of the underlying master equation: 
\begin{equation}
\partial _{t}P(\{\sigma _{{\bf r}}\},t)=\sum_{\{\sigma _{{\bf r}}^{\prime
}\}}\left\{ W[\{\sigma _{{\bf r}}^{\prime }\}|\{\sigma _{{\bf r}%
}\}]P(\{\sigma _{{\bf r}}^{\prime }\},t)-W[\{\sigma _{{\bf r}}\}|\{\sigma _{%
{\bf r}}^{\prime }\}]P(\{\sigma _{{\bf r}}\},t)\right\} \text{ .}
\label{master}
\end{equation}
Here, $W[\{\sigma _{{\bf r}}\}|\{\sigma _{{\bf r}}^{\prime }\}]$ denotes the
transition rate, per unit time, from configuration $\{\sigma _{{\bf r}}\}$
into a new configuration $\{\sigma _{{\bf r}}^{\prime }\}$ which may differ
only by one vacancy hop. In discrete time, it is just the Metropolis rate
specified above. Even though Eqn (\ref{master}) is only linear in the
probabilities, a complete solution is usually feasible only for systems
which are either very small ($L\lesssim 3$) or restricted to one dimension.
Both cases are only of limited interest here. However, it is often possible
to determine a particular eigenvector, namely the one associated with the
(non-degenerate) eigenvalue zero: this provides us with the {\em stationary
limits}, $P_{o}(\{\sigma _{{\bf r}}\})\equiv \lim_{t\rightarrow \infty
}P(\{\sigma _{{\bf r}}\},t)$, of the full distributions.

This procedure is particularly simple for the case of brick wall
(reflecting) boundary conditions. Here, we can immediately write down the
internal energy (the ``Hamiltonian'') of the system, being the electrostatic
energy of a single charge in a uniform electric field. Thus, $P_{o}(\{\sigma
_{{\bf r}}\})\propto \exp \left[ \sum_{{\bf r}}Eay\delta _{{\sigma _{{\bf r}%
},}0}\right] $ is just the associated equilibrium Boltzmann factor. Since
interparticle interactions are restricted to the excluded volume constraint, 
$P_{o}$ is ``color-blind'', i.e., it gives equal weight to all
configurations of {\em particles} at {\em fixed} vacancy position. Of
course, any additional interactions could easily be incorporated into the
Hamiltonian.

In contrast, the toroidal geometry of periodic boundary conditions, in
combination with a uniform drive, prevents the existence of a global, {\em %
time-independent} potential. Therefore, the steady state is far from
equilibrium. Yet fortunately, the stationary distribution is still exactly
known \cite{Spitzer} and even simpler: $P_{o}$ is {\em completely} uniform,
giving equal weight to {\em all} configurations of vacancy {\em and}
particles. In this case, however, it is entirely unknown how to generalize
this solution to more complex interparticle interactions.

Given the stationary distributions, the steady-state {\em densities},
defined via $\phi _{o}({\bf r)}\equiv \lim_{t\rightarrow \infty }\phi ({\bf r%
},t)$ and $\psi _{o}({\bf r)}\equiv \lim_{t\rightarrow \infty }\psi ({\bf r}%
,t)$, are easily computed. For reflecting boundary conditions, $\phi _{o}$
is spatially {\em inhomogeneous}, following the usual exponential profile.
Since the steady state is color-blind, the magnetization density is uniform.
For simplicity, we place the coordinate origin into the center of the
system. Thus,

\begin{equation}
\phi _{o}({\bf r)}=C\exp (Eay)\text{ \quad and\quad }\psi _{o}({\bf r)}=0%
\text{ ,}  \label{exBW}
\end{equation}
where the normalization $C=\sinh (Ea/2)/[L^{d-1}\sinh (Ea(L+1)/2)]$ ensures
that the system contains only one vacancy. The second equation just reflects
the fact that there are equal numbers of positive and negative particles 
\cite{o+e}. It is noteworthy, and may not be entirely trivial at first
sight, that the inhomogeneities in $\phi $ leave no trace at all in the
magnetization distribution.

The corresponding results for periodic boundary conditions are even simpler: 
\begin{equation}
\phi _{o}({\bf r)}=\frac{1}{L^{d}}\text{ \quad and\quad }\psi _{o}({\bf r)}=0%
\text{ .}  \label{exP}
\end{equation}
Here, $\phi _{o}({\bf r)}$ is also uniform, reflecting a single vacancy in a
system of $L^{d}$ sites.

Finally, the saturation value for the number of broken bonds is readily
found, since the particle configurations in the steady state are completely
random. Since each of the four types of particle-particle bonds is equally
likely, to leading order in $1/L$, we obtain 
\[
{\cal A}_{sat}(L,E)\equiv \lim_{t\rightarrow \infty }{\cal A}(L,E;t)=\frac{d%
}{2}L^{d}\left[ 1+O(L^{-1})\right] \text{ .} 
\]
The corrections include surface terms which distinguish the boundary
conditions: for BWBC, we find $O(L^{-1})=-2^{d+1}/(dL)$ while the leading
correction for PBC is only $1/L^{d}$. The vacancy plays an even smaller role
since it affects only $2d$ bonds: this contribution is neglected here.

Having established the key features of the steady state, we now turn to
simulation results. 

\section{Separation of time scales. \label{sec4}}

Our aim in this section is to motivate the key ingredient which will be
required for most of our mean-field results, namely, that the vacancy
equilibrates on a much faster time scale than the particles. Since this
observation will be based on Monte Carlo data, we first summarize the
parameters of our numerical work.

We have performed detailed simulations of our model, using two-dimensional
square lattices with $L$ ranging from $20$ to $60$. The lattice constant $a$
is set to $1$. Both types of boundary conditions, reflecting and periodic,
have been implemented. For reflecting boundary conditions, the external
field varies between $0$ and $1.0$. Larger values of $E$ or $L$ equilibrate
too slowly for our computational resources. For periodic boundary
conditions, equilibration takes place much faster, and therefore, $E$ varies
between $0$ and $10.0$. Our data are averaged over $10^{2}-10^{5}$
realizations, depending on the desired accuracy. The length of the runs
varies with system size, up to a maximum of $10^{8}$ MCS.

\begin{figure}[tbp]
 \input epsf
  \hfill\hfill
\begin{minipage}{0.16\textwidth}
  \epsfxsize = \textwidth \epsfysize = 1.5\textwidth \hfill
  \epsfbox{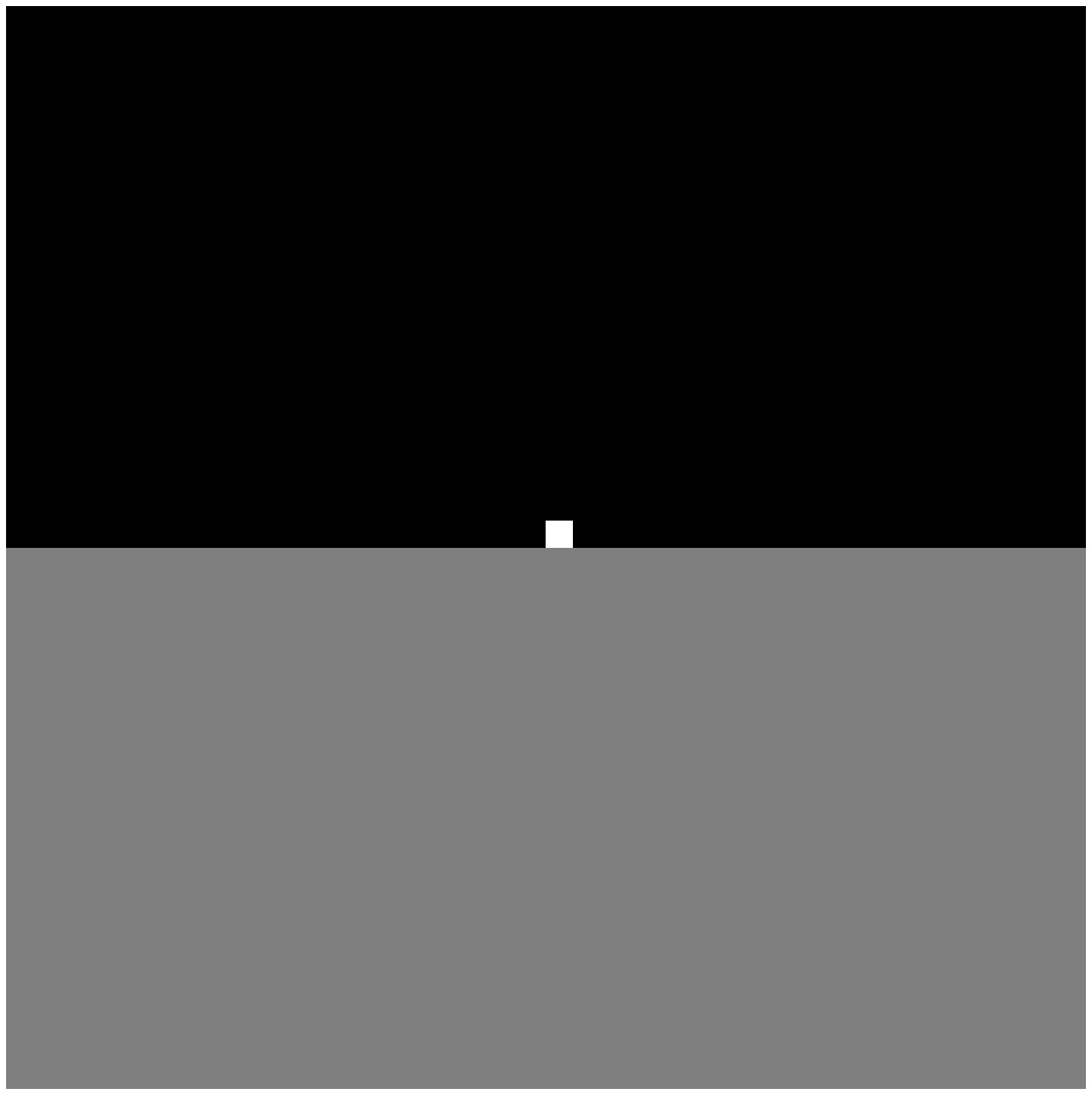} \hfill
    \begin{center} $0$ MCS \end{center}
  \end{minipage}
  \hfill \hfill
\begin{minipage}{0.16\textwidth}
  \epsfxsize = \textwidth \epsfysize = 1.5\textwidth \hfill
  \epsfbox{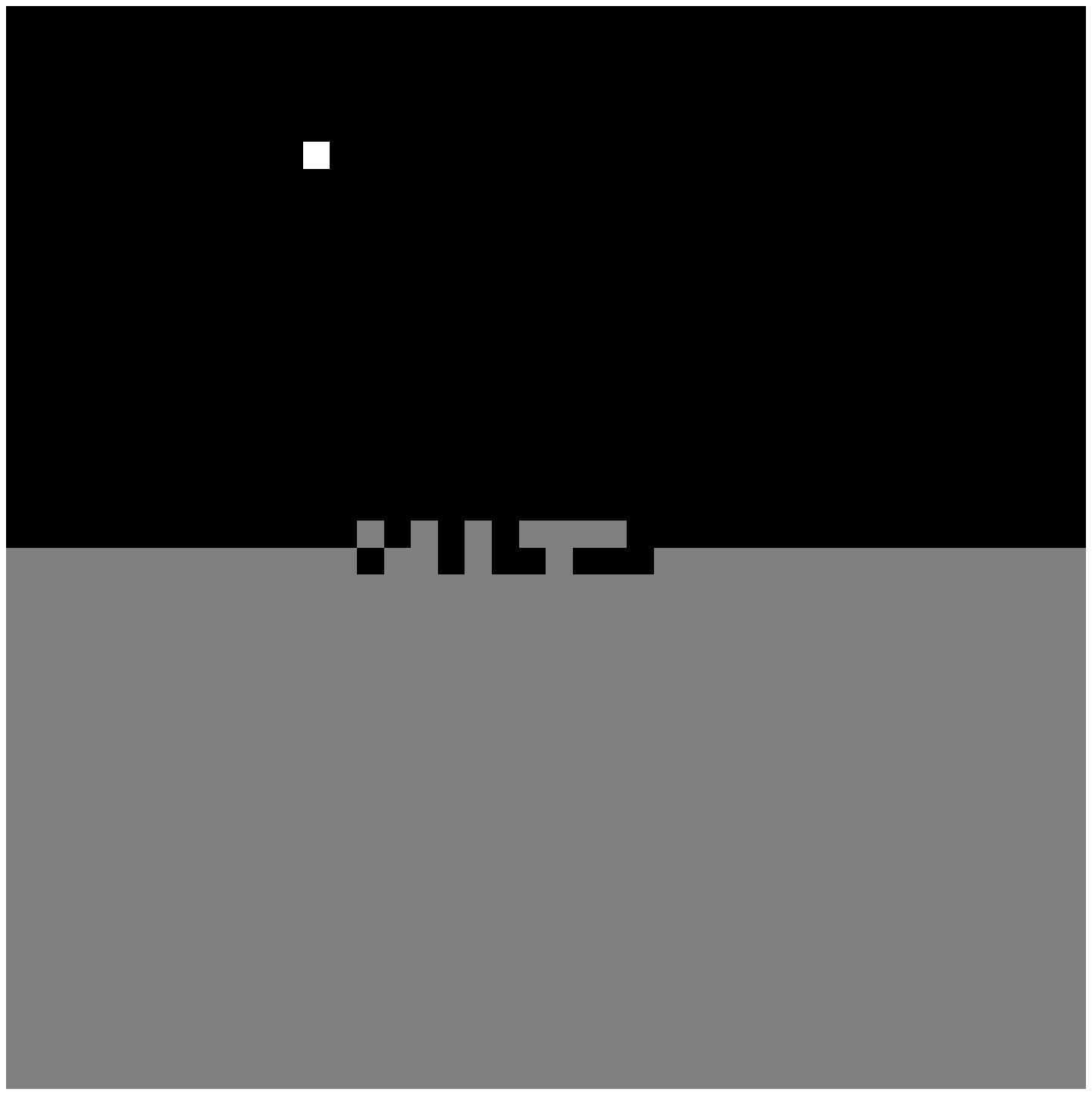} \hfill
    \begin{center} $10^{3}$ MCS \end{center}
  \end{minipage}
  \hfill \hfill
\begin{minipage}{0.16\textwidth}
  \epsfxsize = \textwidth \epsfysize = 1.5\textwidth \hfill
  \epsfbox{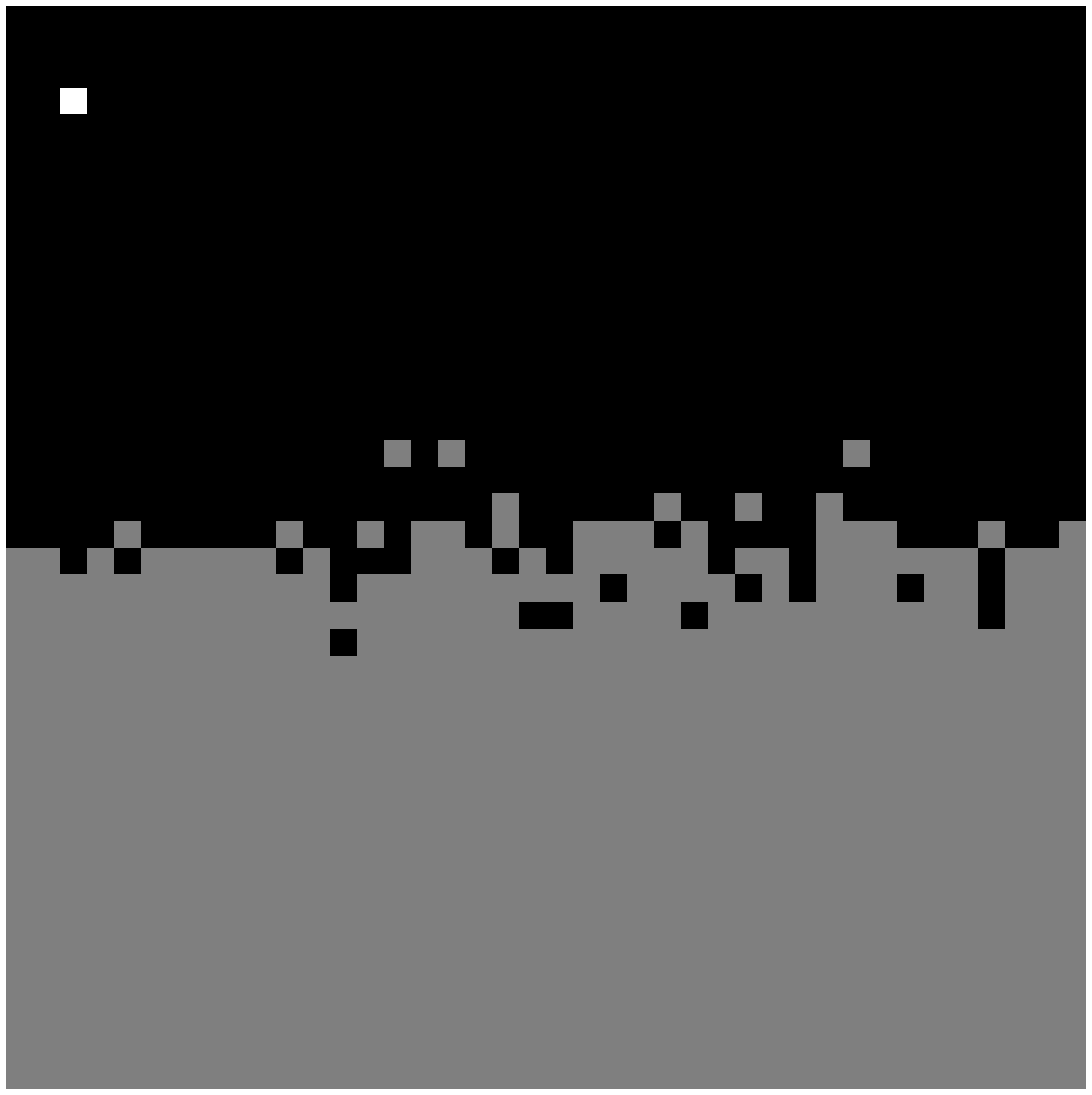} \hfill
    \begin{center}  $10^{4}$ MCS \end{center}
  \end{minipage}
  \hfill\hfill \vspace{0.02\textwidth}
\par
\hfill\hfill
\begin{minipage}{0.16\textwidth}
  \epsfxsize = \textwidth \epsfysize = 1.5\textwidth \hfill
  \epsfbox{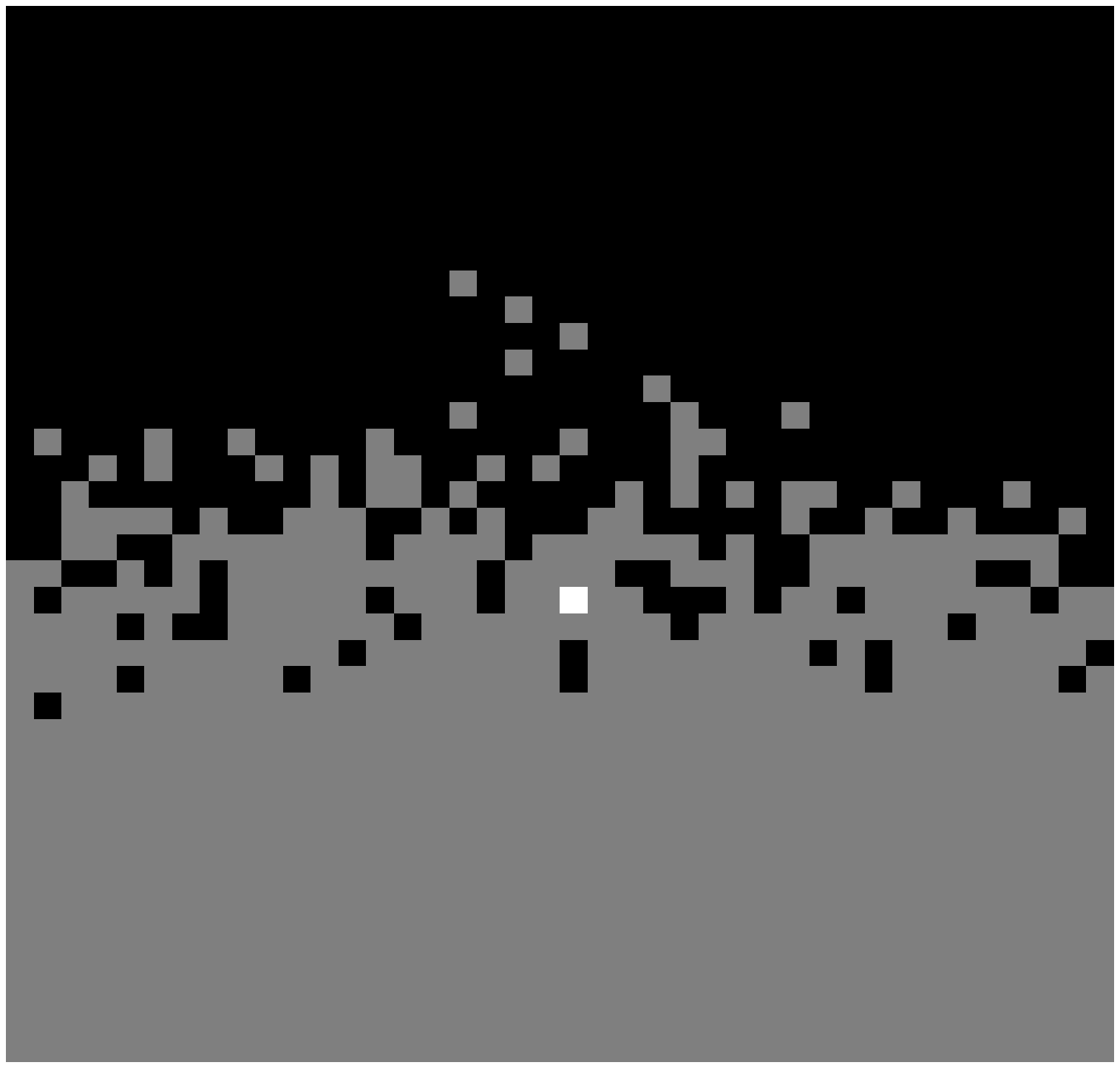}
    \begin{center} $10^{5}$ MCS  \end{center}
  \end{minipage}
  \hfill \hfill
\begin{minipage}{0.16\textwidth}
  \epsfxsize = \textwidth \epsfysize = 1.5\textwidth \hfill
  \epsfbox{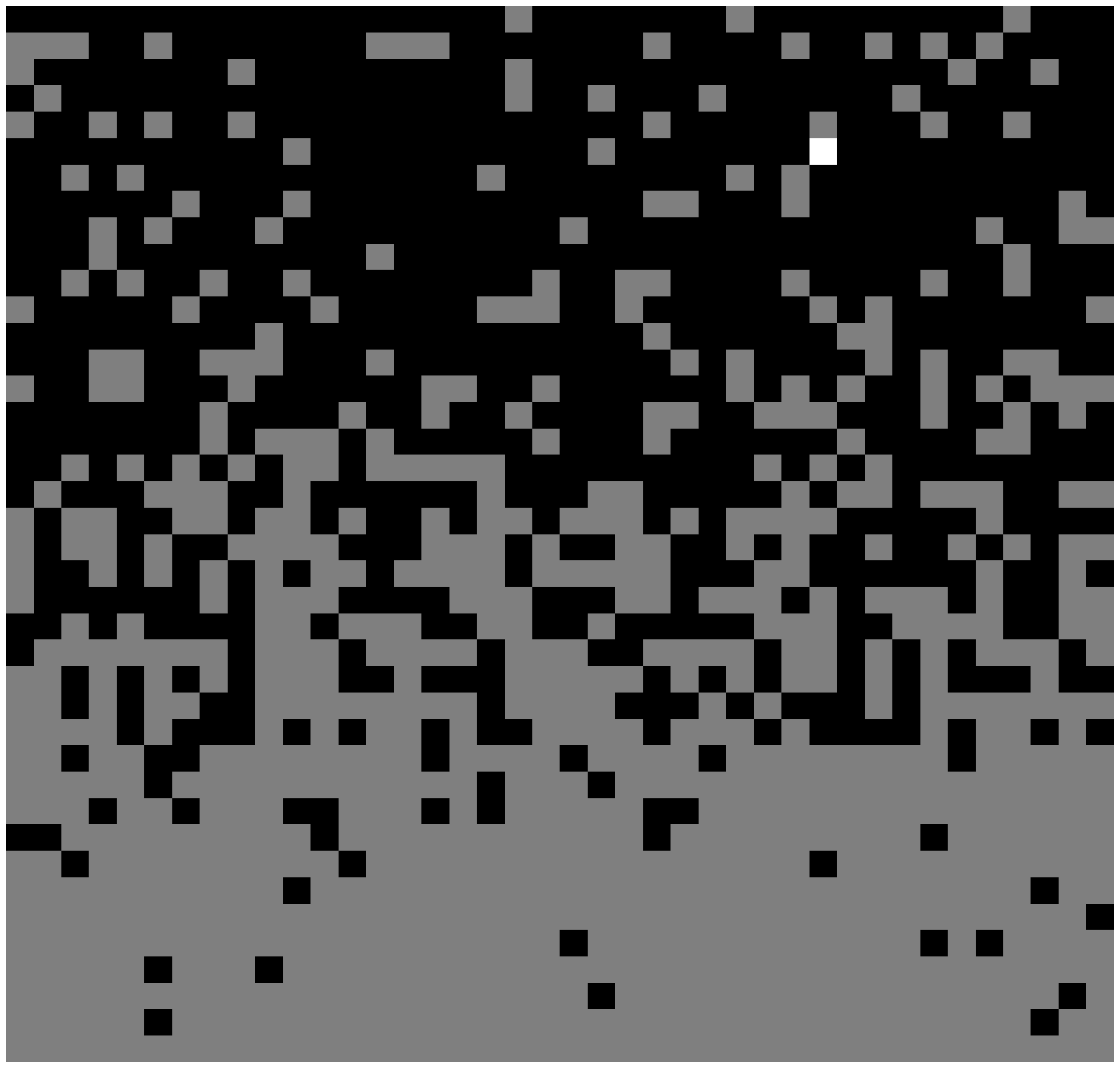}
    \begin{center}   $10^{6}$ MCS\end{center}
  \end{minipage}
  \hfill \hfill
\begin{minipage}{0.16\textwidth}
  \epsfxsize = \textwidth \epsfysize = 1.5\textwidth \hfill
  \epsfbox{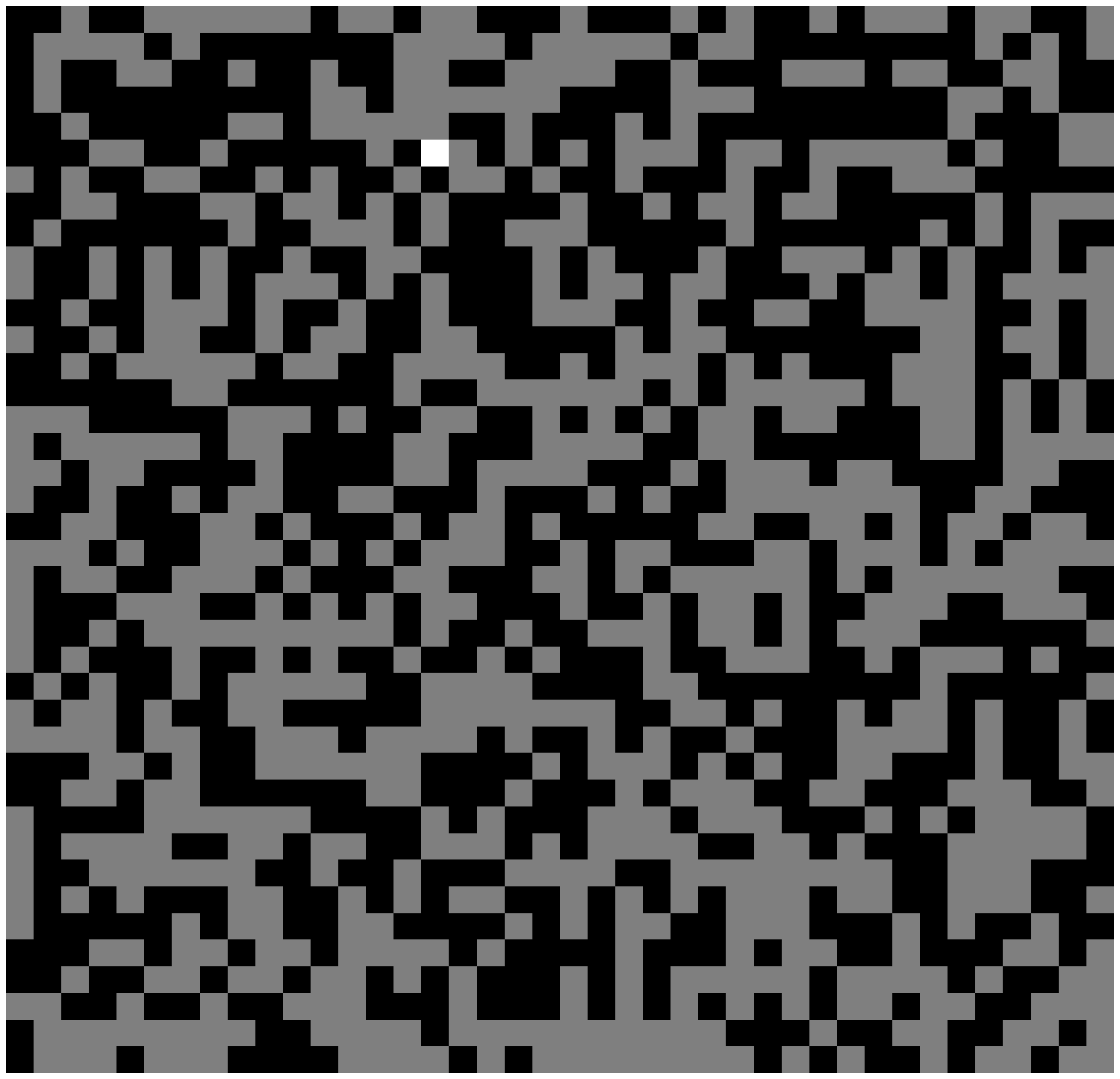}
    \begin{center}  $10^{7}$ MCS \end{center}
  \end{minipage}
  \hfill \hfill
\vspace{0.03\textwidth}
\caption{Sequence of snapshots showing the disordering process of $40x40$
system  with $E= 0.05$ and BWBC. The black and gray squares represent the 
two types of particles($\sigma = \pm 1$) and the white square denotes the
vacancy($\sigma = 0$).  The configurations were recorded after  $0,10^3,
10^4,10^5,10^6 \ $and$ \ 10^7$ MCS.  }
\label{6latticepic}
\end{figure}
\noindent

To obtain a visual impression of the disordering process, Fig.~1 shows the
evolution of a typical configuration, in a $40\times 40$ lattice with $%
E=0.05 $ and brickwall boundary conditions. The initial interface,
completely smooth at $t=0$, begins to break up as increasing numbers of
particles are transported into the oppositely colored half of the system.
Eventually, the system becomes completely disordered. Clearly, the number of
broken bonds, ${\cal A}(L,E;t)$, is a suitable quantitative measure for the
growing disorder, shown in Fig.~2 for the same set of system parameters.

\begin{figure}[tbp]
\input epsf
\begin{center}
\begin{minipage}{0.45\textwidth}
  \epsfxsize = \textwidth \epsfysize = .85\textwidth \hfill
  \epsfbox{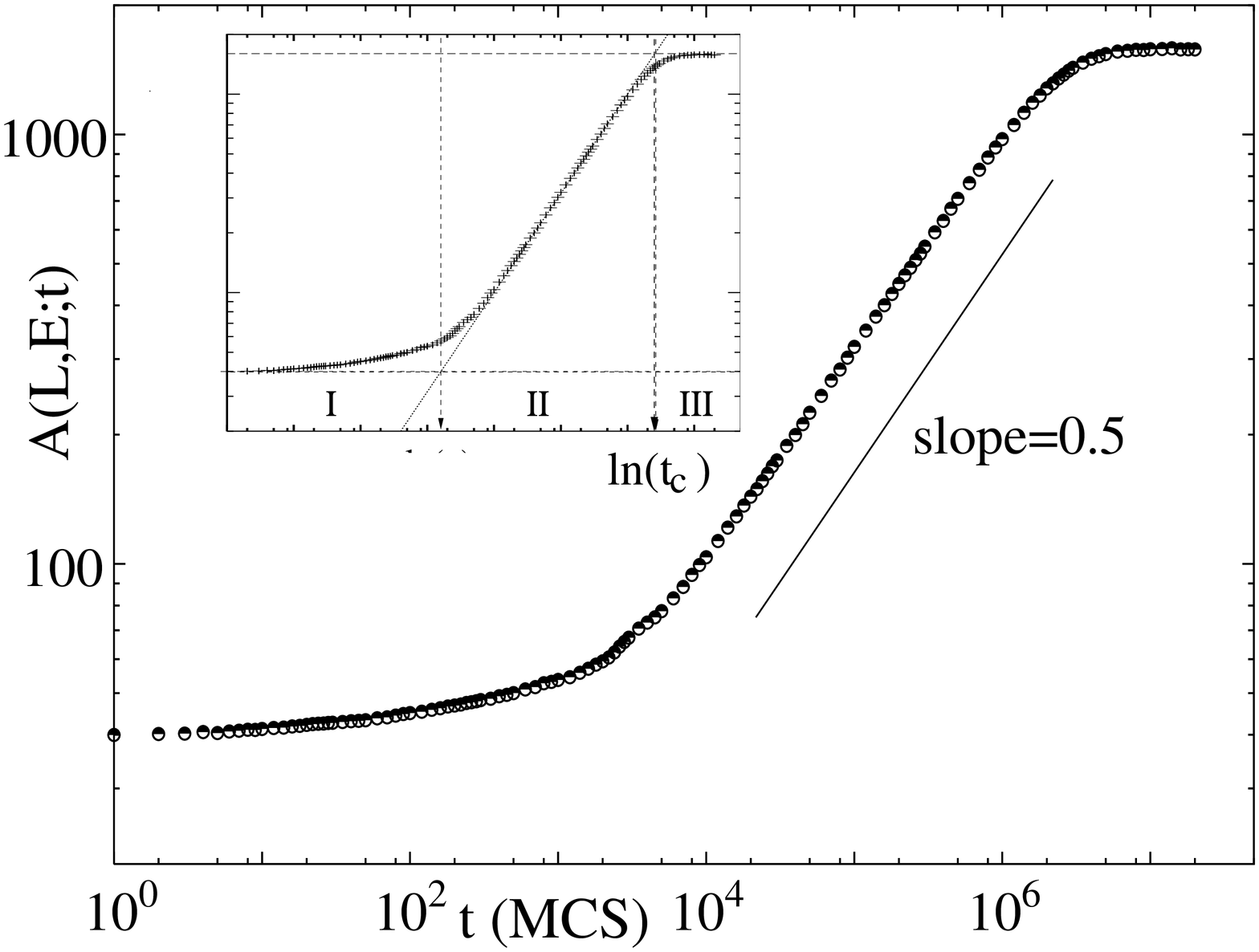}
    \vspace{-.2cm}
\end{minipage}
\end{center}
\caption{Plot of $A(L,E;t)$ vs  $ t$, for $L=40$, $E=0.05$ and BWBC. 
The inset shows the emergence  of an {\em early }regime (I), an  
{\em intermediate}, or {\em scaling}, regime (II), and a {\em late }or
{\em saturation }regime (III). See text for additional details.}
\label{p2_3regimes}
\end{figure}
\noindent

Similar to the unbiased case \cite{TKSZ}, it clearly exhibits three regimes,
drawn schematically in the inset: an {\em early} regime (I), an {\em %
intermediate} regime (II), and a {\em late} or {\em saturation }regime (III)
in which the system has reached steady state. As the system size increases,
the intermediate regime spans a widening time range, suggesting that the
three regimes are temporally well separated. In practice, this is already
the case for $L\gtrsim 20$. This separation of time scales is also observed
for larger values of the bias. Physically, the three regimes are
characterized as follows. For {\em early} times (regime I), the vacancy is
still localized in the vicinity of its starting point. Given the properties
of random walks, this regime is expected to depend strongly on the
dimensionality. After a time of $O(L^{2}$), however, the vacancy has
explored the whole system and is effectively equilibrated. This marks the
onset of the {\em intermediate} regime. In contrast, the particle
distribution is still strongly inhomogeneous, equilibrating only upon
entering the {\em saturation} regime. A completely analogous picture emerges
for periodic boundary conditions.

Our goal in the following is to test for dynamic scaling. Since our data
involve the three variables $L$, $E$ and $t$, as well as two types of
boundary conditions, the appropriate scaling forms may not be a priori
obvious. We shall therefore first consider an analytic approach, before
turning to a detailed comparison with the data.

\section{Mean-field theory}

\subsection{The equations of motion.}

Since an exact solution of the full master equation is not available, we
seek a simpler formulation of the dynamics. For our purposes, a set of
equations of motion for the two conserved densities, $\phi $ and $\psi $,
provides a suitable starting point. We immediately focus on the $d$%
-dimensional case. A {\em spatially} {\em discrete }version of the desired
equations is easily obtained from the master equation, via the definition (%
\ref{dens}). Not surprisingly, the excluded volume interaction mixes
correlation functions of different orders, generating the usual BBGKY
hierarchy. To obtain a closed set of equations, a {\em mean-field}
approximation is required: higher correlations are approximated by products
of appropriate single-point averages. Finally, we take the continuum limit,
by letting the lattice constant $a$ vanish at fixed {\em physical} system
size $L$ and field $E$. The microscopic time scale is identified with $%
a^{2}/2d$. Thus, the resulting densities are functions of a {\em continuous} 
$d$-dimensional coordinate ${\bf r}\equiv (x_{1},\ldots ,x_{d-1},x_{d}\equiv
y)$ and time $t$. The spatial origin is chosen at the center of the system
so that $-L/2\leq x_{i}\leq L/2$ for $i=1,2,...d$. Letting ${\bf \hat{y}}$
denote the unit vector along the direction of $E$, we finally obtain the
basic equations of motion: 
\begin{eqnarray}
\partial _{t}\phi ({\bf r},t) &=&{\bf \nabla }[{\bf \nabla }\phi ({\bf r}%
,t)-E{\bf \hat{y}}\phi ({\bf r},t)]  \label{hole} \\
\partial _{t}\psi ({\bf r},t) &=&{\bf \nabla }[\phi ({\bf r},t){\bf \nabla }%
\psi ({\bf r},t)-\psi ({\bf r},t){\bf \nabla }\phi ({\bf r},t)+E{\bf \hat{y}}%
\phi ({\bf r},t)\psi ({\bf r},t)]  \label{particle}
\end{eqnarray}
Here, ${\bf \nabla }$ is the $d$-dimensional gradient. Clearly, both
equations take the form of continuity equations, reflecting the conservation
laws. The right-hand sides are the (negative) gradients of the associated
hole and magnetization currents. In the field-free limit $E=0$, these are
just the results of Ref. \cite{TKSZ}. Of course, the bias induces additional
terms which reflect systematic Ohmic currents: Since particles can only move
when the vacancy is present, the extra {\em magnetization} current, $-E\hat{y%
}\phi \psi $, is proportional to $\phi $. In contrast, the {\em hole}
current $E\hat{y}\phi $ is independent of the local particle background, as
one should have expected. We also note here that $E$ has dimensions of
inverse length. This suggests that the {\em dimensionless} combination $LE$
will emerge as a convenient scaling variable.

Eqns (\ref{hole}) and (\ref{particle}) have to be supplemented with the
constraints on the total hole and particle \cite{o+e} numbers: 
\begin{equation}
\int_{V}\phi ({\bf r},t)=1\quad \text{and}\quad \int_{V}\psi ({\bf r},t)=0%
\text{.}  \label{norm}
\end{equation}
Here, $V=L^{d}$ is the volume of the system. Next, we consider the boundary
conditions. In the PBC case, the solutions of Eqns (\ref{hole}) and (\ref
{particle}) must be fully periodic functions of ${\bf r}$, with period $L$.
For reflecting boundary conditions, the periodicity of the solutions is
restricted to the transverse subspace. {\em Along} $E$, we demand instead
that the hole and magnetization {\em currents} vanish on the boundary $y=\pm
L/2$.

Finally, we specify the initial conditions. The vacancy starts at the origin 
\begin{equation}
\phi ({\bf r},0)=\delta ({\bf r})  \label{iniphi}
\end{equation}
and the particles are perfectly phase-separated: 
\begin{equation}
\psi ({\bf r},0)=2\theta (y)-1  \label{inipsi}
\end{equation}
This completes the discussion of the equations of motion and their
constraints. We now focus on their solution.

\subsection{Solutions.}

As a starting point, we first establish the steady-state solutions of Eqns (%
\ref{hole}) and (\ref{particle}). When turning to the full time dependence,
progress is only possible if we simplify the equations of motion by invoking
the separation of time scales. We shall mostly focus on {\em brick wall}
boundary conditions, since this situation is physically more complex, as we
shall shortly see. At the end, we briefly summarize our results for the
fully periodic case.

\begin{figure}[tbp]
\input epsf
\begin{center}
\begin{minipage}{0.45\textwidth}
  \epsfxsize = \textwidth \epsfysize = .75\textwidth \hfill
  \epsfbox{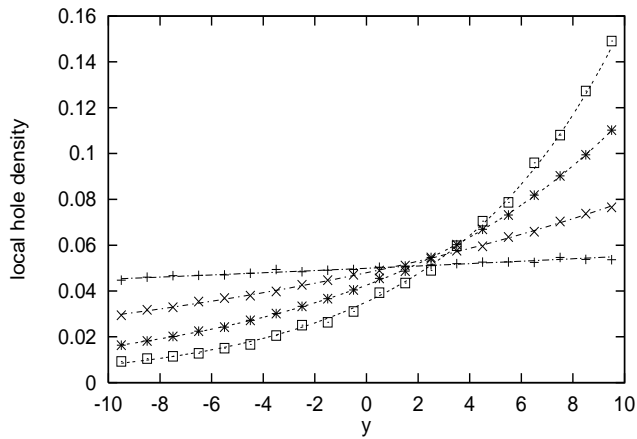}
    \vspace{-.2cm}
\end{minipage}
\end{center}
\caption{Simulation data for the steady state hole profile 
with  $L=20$ and  $E = 0.01(+), 
0.05(\times), 0.10(\ast)$ and $ 0.15(\Box)$, \ BWBC.
The dashed lines denote the mean-field profiles,Eqn({\ref{phio}}) 
and (\ref{C}),  at the same parameter values.}
\label{p3_hole}
\end{figure}

Anticipating inhomogeneities along the $y$-direction only, due to the bias,
we seek {\em steady-state} solutions of the form $\phi ({\bf r},t)=\phi
_{o}(y)$ and $\psi ({\bf r},t)=\psi _{o}(y)$. Eqn (\ref{hole}) can now be
integrated once, and the reflecting boundary conditions force the
integration constant, i.e., the hole current, to zero. We can immediately
integrate again, whence 
\begin{equation}
\phi _{o}(y)=C\exp (Ey)\text{ .}  \label{phio}
\end{equation}
The constant $C$ follows from the normalization condition, (\ref{norm}): 
\begin{equation}
C=\frac{EL}{2L^{d}\sinh (EL/2)}\text{ .}  \label{C}
\end{equation}
Not surprisingly, this expression is just the continuum limit of the exact
result, Eqn (\ref{exBW}). One should note the emergence of the conjectured
scaling variable $LE$. In Fig.~3, we compare the mean-field profiles for $%
L=20$ and a range of $E$'s with the corresponding Monte Carlo data. The
agreement is clearly excellent. Finally, a similar integration of Eqn (\ref
{particle}), using (\ref{phio}), yields $\psi _{o}(y)=0$, also consistent
with the exact form.

Unfortunately, the solution of the time-dependent equations is not so
simple. The full solution $\phi ({\bf r},t)$ of Eqn (\ref{hole}) must be
found and inserted into Eqn (\ref{particle}). The latter poses a formidable
problem: even though it is a linear, second order partial differential
equation of parabolic type, its coefficients are rather complicated
functions of space and time. Fortunately, a complete solution is possible if
we restrict our attention to the {\em intermediate and late} regimes. Here,
the vacancy density has already reached steady state, so that only the
purely exponential $\phi _{o}(y)$ enters into Eqn (\ref{particle}). Of
course, the reduced equation should be supplemented with the initial
condition (\ref{inipsi}). Symmetry considerations as well as the data
suggest that the magnetization density is uniform transverse to $E$, so that
we seek a solution of the form $\psi ({\bf r},t)=\psi (y,t)$. Letting $%
\partial $ denote the partial derivative with respect to $y$, we arrive at a
much simpler version of Eqn (\ref{particle}), describing the magnetization
density at intermediate and late times: 
\begin{equation}
\partial _{t}\psi (y,t)=\partial [\phi _{o}(y)\partial \psi (y,t)]\text{ .}
\label{latepsi}
\end{equation}
It is subject to the initial condition (\ref{inipsi}) and the brick wall
boundary condition 
\begin{equation}
\partial \psi (\pm \frac{L}{2},t)=0  \label{bc1}
\end{equation}
which ensures that the magnetization current through the ends of the system
vanishes.

This differential equation can be reduced to an eigenvalue problem with a
complete and orthonormal set of eigenfunctions $\{U_{n}(y)\}$ which are
linear combinations of Bessel functions. The eigenvalues $\kappa _{n}$ are
real, discrete and strictly positive. Deferring all mathematical details to
the Appendix, we only quote the form of the solution, expressed as an
eigenfunction expansion:
\begin{equation}
\psi (y,t)=\sum\limits_{n=1}^{\infty }A_{n}U_{n}(y)\exp (-\kappa _{n}t)
\label{psisol}
\end{equation}
The expansion coefficients $A_{n}$ are chosen so as to match the initial
condition (\ref{inipsi}).

\begin{figure}[tbp]
\input epsf
\begin{center}
\begin{minipage}{0.45\textwidth}
  \epsfxsize = \textwidth \epsfysize = .75\textwidth \hfill
  \epsfbox{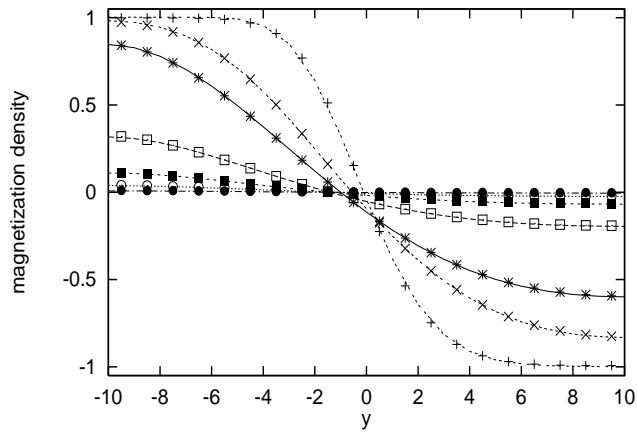}
    \vspace{-.2cm}
\end{minipage}
\end{center}
\caption{Simulation data for charge density profile with $L=20$,
$E=0.1$ and BWBC after $10^{4}(+),5x10^{4}(\times),10^{5}(\ast),
3x10^{5}(\Box), \\  5x10^{5}( \blacksquare),7x10^{5}(\circ)$,
and $10^{6}(\bullet)$ MCS. The dashed lines show the corresponding
mean-field profiles based on Eqn(\ref{psisol}).  See text for details.}
\label{p4_charge}
\end{figure}

\noindent
Combined with the explicit expressions for its
constituents (cf. the Appendix), Eqn (\ref{psisol}) is exact within our
mean-field approach. It agrees remarkably well with Monte Carlo data, as
demonstrated by Fig.~4. 
There, we compare the sum of the first $1000$ terms
in Eqn (\ref{psisol}), for several values of $t$, with simulation results. A
single fit parameter is required, which converts $t$ into the number of MCS.
One sees clearly how the initially phase-segregated system disorders with
time.

Having established the structure of the densities, we turn to the disorder
parameter. Being the number of broken bonds, it is directly related to the
(nearest-neighbor) Ising ``energy'' of our model, $\left\langle -\sum_{<{\bf %
r},{\bf r}^{\prime }>}\sigma _{{\bf r}}\sigma _{{\bf r}^{\prime
}}\right\rangle $. Since the continuum limit of the latter is proportional
to $\int_{V}\psi ^{2}$ \cite{Amit}, we obtain \cite{TKSZ}: 
\begin{equation}
{\cal A}(L,E;t)=\frac{d}{2}L^{d}\left[ 1-\frac{1}{L}\int_{-L/2}^{+L/2}dy\,%
\psi ^{2}(y,t)\right] +O(L^{d-1})\text{ .}  \label{bond1}
\end{equation}
We emphasize that the boundary conditions only affect surface terms. These
will be neglected in the following, leading only to small errors since $%
L\geq 20$ in the data. To the same accuracy, the initial condition for $%
{\cal A}$ is just ${\cal A}(L,E;0)$ $=0$. Using Eqn\ (\ref{psisol}) and
carrying out the integral, we find the time evolution of the disorder
parameter, for {\em brick wall} boundary conditions: 
\begin{equation}
{\cal A}_{R}(L,E;t)=\frac{d}{2}L^{d}\left[ 1-\frac{1}{EL}\sum\limits_{n=1}^{%
\infty }A_{n}^{2}\exp (-2\kappa _{n}t)\right] \text{ }.  \label{bond2}
\end{equation}
Here, we have introduced a subscript, $R$ (``reflecting'') or $P$
(``periodic''), to distinguish the two types of boundary conditions.

We conclude by focusing briefly on the case of {\em fully periodic} boundary
conditions. The steady-state solutions of Eqns (\ref{hole}) and (\ref
{particle}), consistent with the constraints (\ref{norm}), are $\phi _{o}(%
{\bf r)}=1/L^{d}$ and $\psi _{o}({\bf r)}=0$. To obtain the time-dependent
excess density $\psi ({\bf r},t{\bf )}$, we invoke the separation of time
scales again. Noting that the vacancy density has equilibrated at the
beginning of the intermediate regime, we insert its steady-state form, $\phi
_{o}$, into Eqn (\ref{particle}), whence 
\[
\partial _{t}\psi ({\bf r},t)={\bf \nabla }[\phi _{o}({\bf \nabla +}E{\bf 
\hat{y}})\psi ({\bf r},t)]
\]
This simplified form holds in the intermediate and late regimes. A
Galilei-transformation ${\bf r}\equiv {\bf r}^{\prime }+E\phi _{o}{\bf \hat{y%
}}t$ recasts it as an ordinary diffusion equation, with diffusion
coefficient $\phi _{o}$: 
\begin{equation}
\partial _{t}\psi ({\bf r}^{\prime },t)=\phi _{o}\nabla ^{2}\psi ({\bf r}%
^{\prime },t)  \label{psilatepb}
\end{equation}
Quite remarkably, we observe that all effects of the field disappear in a
suitably chosen {\em co-moving} frame. Thus, the solution of Eqn (\ref
{psilatepb}) and the accompanying disorder parameter can be read off
immediately \cite{note} from Ref. \cite{TKSZ}. In the original frame, the
whole profile $\psi ({\bf r},t)$ {\em drifts} with velocity $-EL^{-d}{\bf 
\hat{y}}$. Of course, no such drift is observed in ${\cal A}_{P}(L,E;t)$, by
virtue of the spatial integration in Eqn (\ref{bond1}): 
\begin{equation}
{\cal A}_{P}(L,E;t)=\frac{d}{2}L^{d}\left[ 1-\frac{8}{\pi ^{2}}%
\sum\limits_{n=1,3,...}^{\infty }\frac{\exp (-2\kappa _{n}t)}{n^{2}}\right] 
\text{ }  \label{bond3}
\end{equation}
where $\kappa _{n}\equiv (2\pi n)^{2}/L^{d+2}$. 

This completes the discussion of our mean-field equations and their
solutions. We now turn to the underlying scaling behavior and compare it to
Monte Carlo data.

\section{Dynamic scaling:\ Analytic predictions and numerical tests.}

In this section, we extract the scaling forms for the number of broken
bonds, ${\cal A}(L,E;t)$. We begin with Eqn (\ref{bond2}) for {\em brick wall%
} boundary conditions. In the Appendix, it is shown that both the
coefficients $A_{n}$ and the eigenvalues $\kappa _{n}$ exhibit
characteristic scaling forms: $A_{n}$ is a function of the new scaling
variable $LE$ alone, while the eigenvalues obey 
\[
\kappa _{n}=L^{-(d+2)}g_{n}(LE)\text{ .}
\]
Therefore, the scaling form of ${\cal A}_{R}(L,E;t)$ is apparent, namely, 
\begin{equation}
{\cal A}_{R}(L,E;t)=L^{d}{\cal F}_{R}\left( LE,t/t_{c}\right) \text{.}
\label{Ascalbw}
\end{equation}
where ${\cal F}_{R}$ is an appropriate scaling function. The temporal scale
factor $t_{c}$ is itself still a function of $LE$ and can be chosen in
different ways. Here, we focus on the crossover from the intermediate to the
saturation regime and define $t_{c}$ as a measure for the {\em late
crossover time}, 
\begin{equation}
t_{c}(L,E)\equiv \kappa _{1}^{-1}=L^{d+2}\tau _{c}(LE)  \label{lct}
\end{equation}
The limits of the scaling function $\tau _{c}\equiv g_{1}^{-1}$ are
discussed in the Appendix: It approaches its field-free limit for {\em %
vanishing} $LE$ and increases exponentially with $LE$ in the {\em opposite}
limit, reflecting the increasing confinement of the vacancy. We already note
one of our key results here, namely the emergence of a {\em new }scaling
variable, $LE$. Its physical origin is clear: it determines how easily the
vacancy can escape from the top ($y=L/2$) edge of the system, where the
field tends to localize it, in order to travel to the center of the system
where most of the disordering is taking place. 

It is interesting to contrast the behavior of the disorder parameter, and
hence the scaling function ${\cal F}_{R}$ for ``late'' and ``early'' times, $%
t/t_{c}\gg 1$ and $t/t_{c}\ll 1$ (but already within the intermediate
regime) respectively. For late times, the disorder parameter has saturated
so that $\lim_{y\rightarrow \infty }{\cal F}_{R}(x,y)=d/2$ independent of $%
x\equiv LE$. The short-time limit requires some care since Eqn (\ref{bond2})
does not converge well there. However, using a Poisson resummation (see the
Appendix for details), we can recast it in an alternate form that converges
rapidly for small $\zeta \propto t/t_{c}$: 
\begin{equation}
{\cal A}_{R}(L,E;t)\simeq \frac{d}{2}L^{d}\frac{8}{\sqrt{\pi }}\sqrt{t/t_{s}}%
\left\{ 1+O\left[ \exp (-\pi ^{2}/\zeta ^{2})\right] \right\} 
\label{Abwshort}
\end{equation}
Thus, we conclude that the disorder parameter increases as a power law, $%
t^{\beta }$, for short times, with an exponent $\beta =1/2$. The {\em %
short-time} scale factor exhibits a similar scaling form as the late
crossover time, 
\begin{equation}
t_{s}(L,E)=L^{d+2}\tau _{s}(LE)\text{ .}  \label{sts}
\end{equation}
but we emphasize that its scaling function $\tau _{s}$ {\em differs} from $%
\tau _{c}$ in its dependence on $LE$. This feature finds its origin in the
short-time limit of the scaling function, $\lim_{y\rightarrow 0}{\cal F}%
_{R}(x,y)={\cal G}_{R}(x)y^{1/2}$, expressed by Eqn (\ref{Abwshort}):\ The
presence of the second argument, $x=LE$, leads to the non-trivial prefactor $%
{\cal G}_{R}(x)$ which translates between the two scaling functions $\tau
_{s}$ and $\tau _{c}$.

Before turning to a comparison with the data, we first contrast Eqns (\ref
{Ascalbw}-\ref{sts}) with the corresponding results for periodic boundary
conditions: 
\begin{equation}
{\cal A}_{P}(L,E;t)=L^{d}{\cal F}_{P}\left( tL^{-(d+2)}\right) \equiv L^{d}%
{\cal F}_{P}(t/t_{c})\text{.}  \label{Ascalp}
\end{equation}
All dependence on $E$ {\em disappears} here: The scaling function ${\cal F}%
_{P}$ can be found in Ref. \cite{TKSZ}, and the late crossover time $t_{c}$
is just $L^{d+2}$. A Poisson resummation yields the short-time behavior $%
{\cal A}_{P}\propto \frac{d}{2}L^{d}\sqrt{t/t_{c}}\left\{ 1+O\left[ \exp
(-\pi ^{2}/\zeta ^{2})\right] \right\} $, for $\zeta \propto t/t_{c}\ll 1$.
Thus, we also find a power law increase here, ${\cal A}_{P}\propto t^{\beta }
$, with the {\em same} exponent $\beta =1/2$. 

These predictions are tested in Figs. 5 and 6, for periodic and reflecting
boundary conditions, respectively.

\begin{figure}[tbp]
 \input epsf
\begin{center}   
\begin{minipage}{0.45\textwidth}
  \epsfxsize = \textwidth \epsfysize = 0.85\textwidth \hfill
  \epsfbox{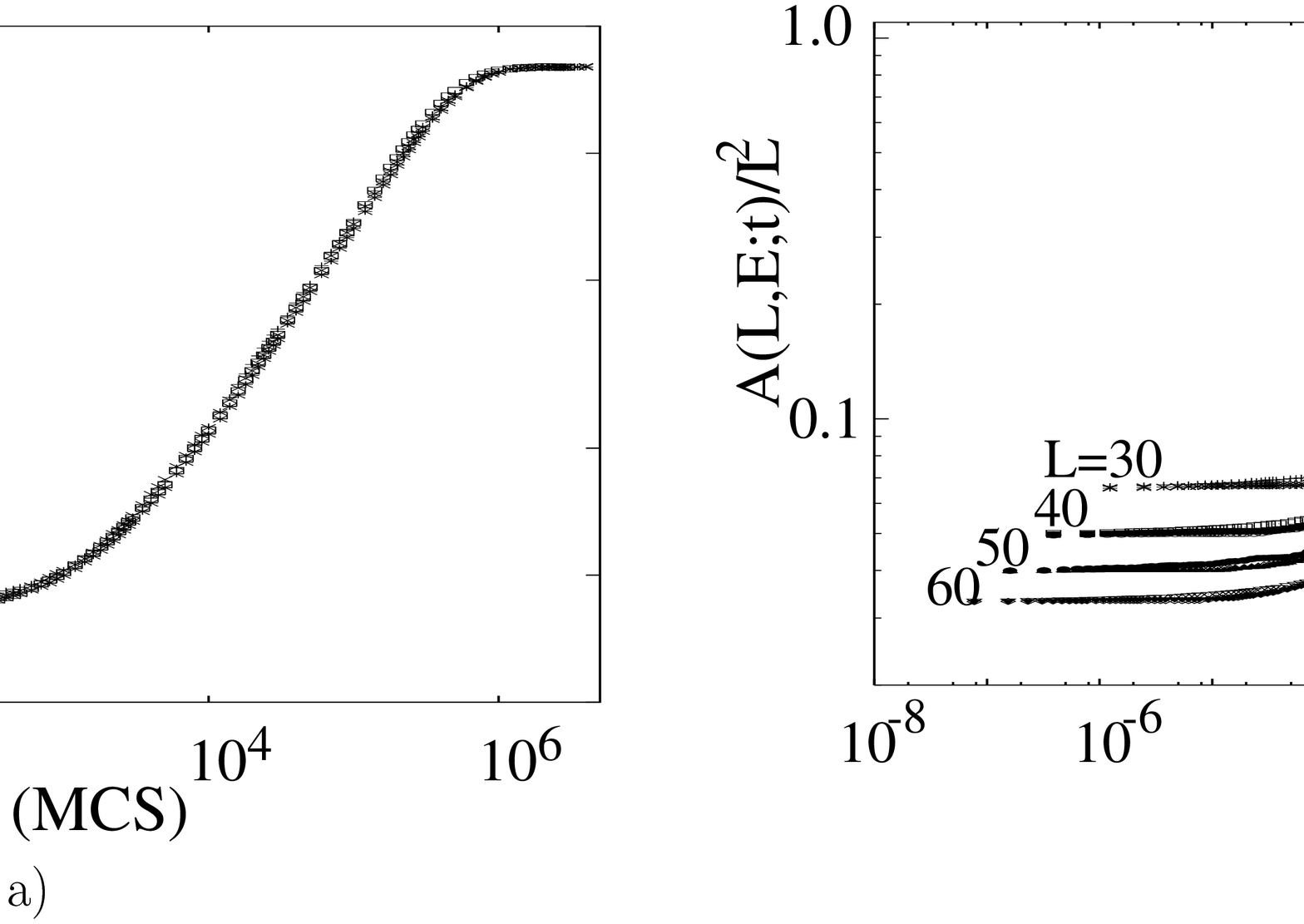}
    \vspace{-1.2cm}
\end{minipage}
\end{center}

\vspace{.4cm}
\caption{Plot of the total number of broken bonds with PBC.
(a) Plot of $A(L,E;t)$ vs $t$ of $40x40$ with $E= 0.00(+), 
0.10(\times), 1.00(\ast)$ and $10.0(\Box)$.  
(b) Scaling plot of $A(L,E;t)/{L^{2}}$ vs \ $t/{L^{4}}$  
for $L=30-60$ with  $E= 0.00, 1.00$ and $10.0$.}
\label{p5ab_BondsPBC}
\end{figure}

\noindent 
In Fig.~5a, we show the disorder
parameter ${\cal A}_{P}$, for {\em one} system size but {\em several} values
of $E$. Some slight deviations are observable at short times, before the
vacancy density equilibrates. In the intermediate and late regimes, however,
it is quite striking, but entirely consistent with our expectations, that
all data points fall onto the same curve, without any scaling being
required. In Fig.~5b, we show a scaling plot for {\em several} $L$ {\em and} 
$E$. Excellent data collapse is observed in the intermediate and late
regimes, where our scaling predictions are expected to hold. Again, we
emphasize that the scaled axes depend only on $L$, but not on $E$.

\begin{figure}[tbp]
 \input epsf

\begin{center}
\begin{minipage}{0.45\textwidth}
  \epsfxsize = \textwidth \epsfysize = 0.85\textwidth \hfill
  \epsfbox{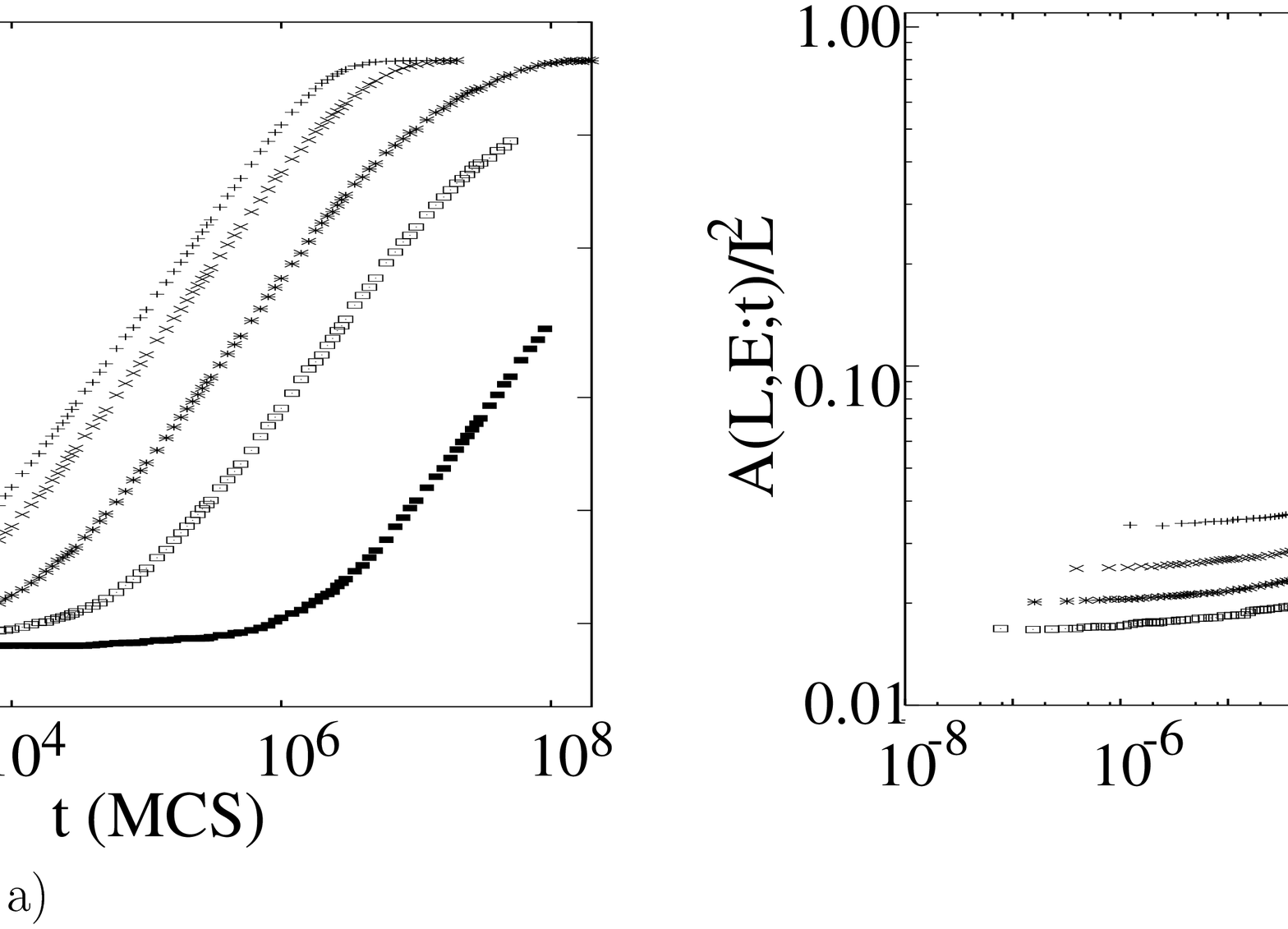}
    \vspace{-1.2cm}
\end{minipage}
\end{center}

\vspace{.4cm}
\caption{Plot of the total number of broken bonds with BWBC. 
(a) Plot of $A(L,E;t)$ vs $ t$ of $40x40$ with 
$E=0.00(+), 0.10(\times), 0.20(\ast), 0.30(\Box)$  and $0.50(\blacksquare)$.  
(b) The scaling plot of $A(L,E;t)/L^{2}$ vs $t/L^{4}$ for L=$30(+),
40(\times), 50(\ast)$ and $60(\Box)$ with $LE= 2.0$. }
\label{p6ab_BondsBWBC}
\end{figure}

\noindent
Fig.~6a shows the disorder parameter ${\cal A}_{R}$ for reflecting boundary
conditions. Several values of $E$ are plotted, for $L=40$. As anticipated,
the mixing process proceeds more slowly for larger values of $E$, since
these tend to confine the vacancy more strongly to the upper edge ($y=+L/2$)
of the system. The corresponding scaling plot is shown in Fig.~6b: Here,
several values of $L$ and $E$ are chosen such that $LE=2.0$ remains {\em %
constant}. The data collapse in the intermediate and late regimes lends full
support to our prediction, Eqn (\ref{Ascalbw}).

Focusing on the intermediate regime ($O(L^{2})\leq t\ll t_{c}$), Figs. 5b
and 6b show very clearly that the disorder parameter increases as a {\em %
power law} there, ${\cal A}(L,E;t)\propto t^{\beta }$. Since this behavior
persists over at least three decades, we can extract a reliable numerical
estimate for the exponent, resulting in $\beta =0.5\pm 0.01$ for both
reflecting and for periodic boundary conditions, in agreement with our
prediction.

Our results so far can be summarized more succinctly. Defining a set of
characteristic exponents and scaling functions, the scaling of the disorder
parameter, for both types of boundary conditions, can by written as 
\begin{equation}
{\cal A}(L,E;t)\sim L^{\alpha }{\cal F}\left( LE,t/t_{c}\right) \qquad \text{%
with\qquad }t_{c}(L,E)\sim L^{z}\tau (LE)\text{ .}  \label{Ascal}
\end{equation}
Remarkably, we find that the {\em exponents} $\alpha $ and $z$ are {\em %
universal}, i.e., independent of boundary conditions and drive. Our results,
$\alpha =d$ which is exact and $z=d+2$ which follows from our mean-field
theory, are completely consistent with the Monte Carlo data and agree with
the corresponding exponents for the unbiased case \cite{TKSZ}. The
short-time behavior can be written as 
\begin{equation}
{\cal A}(L,E;t)\sim L^{\sigma }(t/\tau _{s})^{\beta }\text{ ,}
\label{Ascals}
\end{equation}
with a ``growth'' exponent $\beta =1/2$ which is also manifestly universal.
The additional exponent $\sigma $ is related to the others via the
consistency condition ${\cal A}(L,E;t_{c})\sim {\cal A}(L,E;t\rightarrow
\infty )$, whence $\sigma =d-z\beta =(d-2)/2$. This scaling law is obviously
satisfied by our results.

In contrast to the exponents, the scaling {\em functions }${\cal F}$ and $%
\tau $ are profoundly affected by the bias and the boundary conditions. For
PBC, we just recover the results of the unbiased case, whereas a nontrivial
dependence on a new scaling variable, $LE$, emerges in the brick wall case.

\section{Conclusions\label{sec5}}

In this paper we have analyzed the vacancy-mediated disordering of a binary
alloy, in response to an upquench from zero to infinite temperature. The
system is placed in a gravitational or electric field, and two types of
boundary conditions, reflecting (``brick wall'') and periodic, are studied.
Starting from a perfectly phase-segregated initial configuration, the
vacancy mediates atom exchanges, leading to fully disordered {\em exactly
known} final states. For brickwall boundary conditions, the final state is
an {\em equilibrium} one, characterized by a Boltzmann distribution and an
exponential hole density profile. In contrast, the steady state established
by periodic boundary conditions is a {\em non-equilibrium} one, with
homogeneous configurational distribution and profiles. Using Monte Carlo
simulations, we monitor the time evolution of the system by measuring local
profiles and a disorder parameter, i.e., the number of broken bonds. For
both types of boundary conditions, three temporal regimes are identified: an
early one, where the vacancy has not reached the boundaries of the system,
an intermediate one in which the vacancy has already equilibrated, but the
particle distributions are still inhomogeneous, and a late regime where the
system has reached steady state. To predict scaling exponents and scaling
functions, we develop a theoretical description in terms of a set of
mean-field equations of motion for the local densities. Invoking the
separation of time scales, the mean-field equations can be solved exactly in
the intermediate and late time regimes, providing us with explicit
expressions for the observables of interest. We find that the disorder
parameter exhibits dynamic scaling and observe excellent agreement of Monte
Carlo data and mean-field predictions. Our key result is a set of {\em %
universal scaling exponents}, independent of bias and boundary conditions.
For example, in all cases the number of broken bonds increases as $t^{1/2}$,
before saturating at a value of $O(L^{d})$. In contrast, the scaling {\em %
functions} for the disorder parameter exhibit a {\em lesser} degree of
universality: While they remain independent of the bias in the case of
periodic boundary conditions, an additional scaling variable, $LE$, must be
taken into account for reflecting boundary conditions. This variable also
controls the shape of the excess density (magnetization) profile. 

So far, we have restricted our attention to systems containing only a single
vacancy. However, in real systems, one should expect that the number of
vacancies, $M$, scales with system size \cite{TKSZ}. For generality, we
introduce the \emph{vacancy number exponent} $\gamma $, with $0\leq \gamma
\leq d$, so that $M\sim L^{\gamma }$. Thus, $\gamma =d-1$ describes a
situation where the vacancies (``defects'') prefer the interfacial region
before the upquench occurs. Our previous results correspond to the case $%
\gamma =0$ but are easily extended. We simply need to modify the
normalization condition for the hole density profile, Eqn (\ref{norm}),
so that the constant $C$, given in Eqn (\ref{C}), becomes $%
C=EL/[2L^{d-\gamma }\sinh (EL/2)]$. As a consequence, the late crossover
time, Eqn (\ref{lct}), now scales as $t_{c}(L,E)=L^{d+2-\gamma }\tau
_{c}(LE)$, so that $z=d+2-\gamma $ and $\sigma =(d+\gamma -2)/2$. Of course,
these exponents are again universal. 

We conclude by noting the different symmetries which characterize the BWBC
and the PBC case, in the {\em presence} of the bias. Periodic boundary
conditions are compatible with translation invariance, but violate the
detailed balance condition: the driving force is not compensated by a
chemical potential gradient. In contrast, detailed balance holds for
reflecting boundary conditions, but translation invariance is broken. It is
quite remarkable that the scaling exponents remain unaffected by such
profound differences in symmetry.   

\section{Acknowledgments}

We would like to thank S. Redner for asking a pertinent question which
led to this work. We are also grateful to R.~K.~P. Zia, T.~J. Newman, M.
Howard and W. Loinaz for helpful discussions. This research is 
supported in part by the National Science
Foundation through grant number DMR-9727574, and by the Deutsche
Forschungsgemeinschaft through grant number Zi290/5-1.

\newpage 

\begin{center} 
\bf {APPENDIX}
\end{center}


\subsection{Full solution for the magnetization density.\label{appa}}

In this appendix, we present the mathematical details of solving Eqn\ (\ref
{latepsi}) for brickwall boundary conditions, restricted to the intermediate
and late stages of the disordering process, i.e., $O(L^{2})\ll t$. Thus, in
the following, ``early'' refers to times at the onset of the intermediate
regime while ``late'' times are deeply within the saturation region. To
recall, we seek a solution, $\psi (y,t)$, to the partial differential
equation 
\begin{equation}
\partial _{t}\psi (y,t)=\partial [\phi _{o}(y)\partial \psi (y,t)]
\label{psiA1}
\end{equation}
where 
\[
\phi _{o}(y)=\frac{EL}{2L^{d}\sinh (EL/2)}\exp (Ey)\ 
\]
subject to the initial and boundary condition $\psi (y,0)=2\theta (y)-1$ and 
$\partial \psi (\pm \frac{L}{2},t)=0$. First, we introduce the new variable 
\begin{equation}
x\equiv \int_{y}^{\infty }dy^{\prime }\frac{1}{\phi _{o}(y^{\prime })}=\frac{%
1}{E\phi _{o}(y)}\text{ ,}  \label{x}
\end{equation}
which is strictly positive. This reduces Eqn\ (\ref{psiA1}) to a diffusion
equation with a spatially varying diffusion coefficient: 
\begin{equation}
\partial _{t}\psi (x,t)=Ex\partial _{x}^{2}\psi (x,t).  \label{psiA2}
\end{equation}
subject to the boundary condition $\partial _{x}\psi (x_{\pm },t)=0$ at $%
x_{\pm }\equiv \left[ E\phi _{o}(\pm L/2)\right] ^{-1}$. We note that the
definition (\ref{x}) implies $0<x_{+}<x_{-}$. Next, we separate variables,
according to the ansatz 
\begin{equation}
\psi (x,t)=T(t)f(x).  \label{sep}
\end{equation}
whence we obtain two ordinary differential equations: 
\begin{eqnarray}
\frac{dT}{dt}+\kappa T &=&0  \label{T} \\
\frac{d^{2}f}{dx^{2}}+\frac{\kappa }{Ex}f &=&0\text{ ,}  \label{f}
\end{eqnarray}
The constant $\kappa $ must be positive in order to be consistent with the
steady state solution, $\lim_{t\rightarrow \infty }\psi (y,t)=\psi _{o}(y)=0$%
. The first of these equations describes a simple exponential decay. The
second constitutes a well-defined Hermitean eigenvalue problem, with
eigenvalues $\kappa $ and eigenfunctions $f$. For convenience, we define $%
\kappa /E\equiv \alpha ^{2}/4$ and transform (\ref{f}) into the differential
equation for the {\em Bessel functions }\cite{A+S}, via $u\equiv \alpha 
\sqrt{x}$. The solutions are 
\begin{eqnarray}
T(t) &=&\exp (-\kappa t)  \label{Tsol} \\
f(x) &=&A\alpha \sqrt{x}J_{1}(\alpha \sqrt{x})+B\alpha \sqrt{x}N_{1}(\alpha 
\sqrt{x}).  \label{fsol}
\end{eqnarray}
Here, $A$ and $B$ are integration constants. We have two boundary
conditions, one at each end of the system. Using the recursion relations for
the Bessel functions, we can eliminate one of the integration constants,
e.g., 
\begin{equation}
B=-A\frac{J_{0}(\alpha \sqrt{x_{-}})}{N_{0}(\alpha \sqrt{x_{-}})}  \label{B}
\end{equation}
and specify the allowed eigenvalues $\alpha _{n}$ as the solutions of the
implicit equation 
\begin{eqnarray}
0 &=&N_{0}(\alpha _{n}\sqrt{x_{-}})J_{0}(\alpha _{n}\sqrt{x_{+}}%
)-N_{0}(\alpha _{n}\sqrt{x_{+}})J_{0}(\alpha _{n}\sqrt{x_{-}})
\label{alphan} \\
&\equiv &N_{0}(\lambda z_{n})J_{0}(z_{n})-N_{0}(z_{n})J_{0}(\lambda z_{n})
\label{alphastand}
\end{eqnarray}
The second line, with $z_{n}\equiv \alpha _{n}\sqrt{x_{+}}$ and $\lambda
\equiv \sqrt{\frac{x_{-}}{x_{+}}}=\exp (LE/2)>1$, is a more standard form of
the eigenvalue equation \cite{A+S}. Since both $J_{0}$ and $N_{0}$
oscillate, this equation has infinitely many solutions. The eigenvalues are
real, non-degenerate and discrete; they increase monotonically with $n$. The
lowest ones are easily determined numerically for different $\lambda $. For
large $n\pi /(\lambda -1)$, there is an asymptotic expansion, 
\begin{equation}
z_{n}=\frac{n\pi }{\lambda -1}\left\{ 1-\frac{1}{8\lambda }\left( \frac{%
\lambda -1}{n\pi }\right) ^{2}+O\left[ \left( \frac{\lambda -1}{n\pi }%
\right) ^{4}\right] \right\}   \label{asympt}
\end{equation}
To find the eigenfunctions $\{U_{n}(x)\}$, we need to normalize the $f$'s.
For convenience, we introduce the auxiliary function 
\begin{equation}
F(s)\equiv s[N_{0}(\alpha _{n}\sqrt{x_{-}})J_{1}(s)-J_{0}(\alpha _{n}\sqrt{%
x_{-}})N_{1}(s)]  \label{F}
\end{equation}
and define the normalization constants 
\begin{equation}
c_{n}\equiv \frac{\alpha _{n}}{\left[ F^{2}(\alpha _{n}\sqrt{x_{-}}%
)-F^{2}(\alpha _{n}\sqrt{x_{+}})\right] ^{1/2}}=\frac{\alpha _{n}}{\left[
F^{2}(\lambda z_{n})-F^{2}(z_{n})\right] ^{1/2}}\text{ .}  \label{Cn}
\end{equation}
Then, the eigenfunctions take the form: 
\begin{eqnarray}
U_{n}(x) &=&\frac{c_{n}}{\alpha _{n}}F(\alpha _{n}\sqrt{x})  \nonumber \\
&=&c_{n}\sqrt{x}\left[ N_{0}(\alpha _{n}\sqrt{x_{-}})J_{1}(\alpha _{n}\sqrt{x%
})-J_{0}(\alpha _{n}\sqrt{x_{-}})N_{1}(\alpha _{n}\sqrt{x})\right] \text{ .}
\label{Un}
\end{eqnarray}
They are real and orthonormal: 
\[
\delta
_{nm}=\int_{x_{+}}^{x_{-}}dx\,x^{-1}U_{n}(x)U_{m}(x)=E\int_{-L/2}^{+L/2}dy%
\,U_{n}(y)U_{m}(y)
\]
where the second equality expresses the orthonormality condition in terms of
the original variable $y$. They form a complete set so that the full
solution for the magnetization density can be written as an expansion: 
\begin{equation}
\psi (x,t)=\sum\limits_{n=1}^{\infty }A_{n}U_{n}(x)\exp (-\kappa _{n}t)\ ,
\label{psiA3}
\end{equation}
where $\kappa _{n}=E\alpha _{n}^{2}/4$, and $x=\left[ E\phi _{o}(y)\right]
^{-1}$. The expansion coefficients $A_{n}$ are chosen such that the initial
condition is satisfied. With $x_{o}\equiv \left[ E\phi _{o}(0)\right] ^{-1}$%
, this yields: 
\begin{eqnarray}
A_{n} &=&\frac{4c_{n}}{\alpha _{n}}\left[ N_{0}(\alpha _{n}\sqrt{x_{o}}%
)J_{0}(\alpha _{n}\sqrt{x_{-}})-N_{0}(\alpha _{n}\sqrt{x_{-}})J_{0}(\alpha
_{n}\sqrt{x_{o}})\right]   \label{An} \\
&=&\frac{4}{\left[ F^{2}(\lambda z_{n})-F^{2}(z_{n})\right] ^{1/2}}\text{ }%
\left[ N_{0}(\sqrt{\lambda }z_{n})J_{0}(\lambda z_{n})-N_{0}(\lambda
z_{n})J_{0}(\sqrt{\lambda }z_{n})\right]   \label{Anstand}
\end{eqnarray}

This completes the solution. Of course, it is given rather implicitly in
terms of the eigenvalues. To make our expressions more transparent, we track
the key system parameters, $L$ and $E$, through these manipulations, in
order to exhibit the scaling properties of the theory.

\subsection{Scaling analysis.}

\subsubsection{Eigenvalues and expansion coefficients.}

We first establish the scaling properties of the eigenvalues, $\kappa _{n}$.
Beginning with Eqn (\ref{alphastand}), we conclude that the $z_{n}$'s are
functions of $\lambda $ alone. Since $\lambda =\exp (LE/2)$, each $z_{n}$
depends only on the scaling parameter $LE$. To obtain a similar conclusion
for $\kappa _{n}=E\alpha _{n}^{2}/4=Ez_{n}^{2}/(4x_{+})$, we recall that 
\[
1/x_{+}=E\phi _{o}(+L/2)=\frac{(EL)^{2}\exp (EL/2)}{2L^{d+1}\sinh (EL/2)} 
\]
Thus, the desired scaling form for the eigenvalues is 
\begin{equation}
\kappa _{n}=L^{-(d+2)}g_{n}(LE)  \label{kappan}
\end{equation}
where the scaling function $g_{n}(x)$, with $x\equiv LE$, is given by 
\begin{equation}
g_{n}(x)\equiv z_{n}^{2}(x)\frac{x^{3}\exp (x/2)}{8\sinh (x/2)}  \label{gna}
\end{equation}
Its limits for small and large argument are easily found from Eqn (\ref
{alphastand}) and the asymptotic form (\ref{asympt}): 
\begin{equation}
\lim_{x\rightarrow 0}g_{n}(x)=(n\pi )^{2}\left[ 1+O(x)\right] \quad \text{and%
}\quad \lim_{x\rightarrow \infty }g_{n}(x)\simeq x_{n}\exp (-x/2)\ \text{,}
\label{gnlimits}
\end{equation}
where $x_{n}$ denotes the $n$-th zero of the Bessel function $J_{0}$.

In particular, we are interested in the {\em late crossover time}, 
Eqn (\ref{lct}), defined
as the inverse of the first eigenvalue, 
\begin{equation}
t_{c}(L,E)\equiv \kappa _{1}^{-1}\equiv L^{d+2}\tau _{c}(LE)\text{ .} 
\label{recall_lct}
\end{equation}
This characteristic time is a measure for when the crossover from the
intermediate to the saturation regime occurs. The behavior of its scaling
function, $\tau _{c}(x)\equiv g_{1}^{-1}(x)$, follows immediately from Eqn (%
\ref{gnlimits}): $\lim_{x\rightarrow 0}\tau _{c}(x)=\pi ^{-2}\left[
1+O(x)\right] $ and $\lim_{x\rightarrow \infty }\tau _{c}(x)\simeq
(0.1729.../x^{3})\exp (x)$.

Finally, we will need the scaling behavior of the expansion coefficients $%
A_{n}$. From Eqn (\ref{Anstand}), it is immediately apparent that these
coefficients depend only on the {\em combination} $LE$ so that $%
A_{n}=A_{n}(LE)$.

\subsubsection{The Poisson resummation of the disorder parameter: short-time
behavior.}

While Eqn (\ref{bond2}) for ${\cal A}(L,E;t)$ is of course completely exact
within our mean-field theory, it converges rapidly only for late times, $%
\kappa _{n}t\gg 1$. There, keeping only the $n=1$ term in the sum already
results in an excellent approximation. In contrast, Eqn (\ref{bond2}) is not
very practical if we wish to extract the observed power law at early times.
Fortunately, a {\em Poisson resummation }\cite{Jones} of Eqn (\ref{bond2})
allows us to recast the disorder parameter in a form that converges rapidly
in the short-time limit. Some details of this procedure form the content of
this section.

We recall Eqn (\ref{bond2}): 
\begin{equation}
{\cal A}_{R}(L,E;t)=\frac{d}{2}L^{d}\left[ 1-\frac{1}{EL}\sum\limits_{n=1}^{%
\infty }A_{n}^{2}\exp (-2\kappa _{n}t)\right] \text{ ,}  \label{recallA}
\end{equation}
where both $A_{n}$ and $\kappa _{n}$ depend on the summation index via the
eigenvalues $z_{n}$. The key to the Poisson resummation resides in the
following three statements:

\noindent \noindent First, from the discussion below Eqn (\ref{bond1}), we
recall the initial condition on the disorder parameter, namely ${\cal A}%
_{R}(L,E;0)=0$. This implies \cite{new} 
\begin{equation}
\frac{1}{EL}\sum\limits_{n=1}^{\infty }A_{n}^{2}=1\text{ ,}  \label{new}
\end{equation}
so t\noindent hat there is no constant term in the short-time expansion of $%
{\cal A}_{R}$.

\noindent Second, considering any {\em finite} number of terms in Eqn 
(\ref{recallA}) can only generate a {\em linear} time dependence, ${\cal A}%
_{R}(L,E;t)\propto t$. Therefore, the {\em anticipated} short-time behavior $%
{\cal A}_{R}\propto \sqrt{t}$ must be controlled by the {\em large} $n$
contributions to the sum. Hence, these are crucial for our purposes.

\noindent \noindent Third, for sufficiently large $n>n_{o}$, we can always
approximate the eigenvalues $z_{n}$ by their explicit asymptotic form, Eqn (%
\ref{asympt}). Since the latter holds provided $n\pi /[\lambda -1]\gg 1$, the
critical $n_{o}\sim \lambda \equiv \exp (LE/2)$ increases rapidly with $LE$.
However, this presents no problem since only finite values of $LE$ are of
interest to us.

In summary, to obtain the short-time behavior of ${\cal A}_{R}$ it is
sufficient to replace the eigenvalues $z_{n}$ by their asymptotic form {\em %
everywhere}. Any errors generated in this way are at most linear in $t$. In
this manner, the dependence on the summation index $n$ becomes explicit, and
we can apply the Poisson resummation formula \cite{Jones}, which holds for
any continuous, bounded function $f(x)$, provided its Fourier transform $%
F(\omega )\equiv 2\int dxf(x)\cos (\omega x)$ is well-defined: 
\begin{equation}
\frac{1}{2}f(0)+\sum\limits_{n=1}^{\infty }f(n\varsigma )=\frac{1}{\varsigma 
}\left\{ \frac{1}{2}F(0)+\sum\limits_{m=1}^{\infty }F(\frac{2\pi m}{%
\varsigma })\right\} \text{ .}  \label{poisson}
\end{equation}
Here, $\varsigma $ is the parameter that controls the convergence of the
sums. In our case, we identify 
\begin{equation}
\varsigma \equiv \sqrt{\frac{4\pi ^{2}(\ln \lambda )^{3}\lambda ^{2}}{%
(\lambda -1)^{3}(\lambda +1)}L^{-(d+2)}t}  \label{zeta}
\end{equation}
The resummation is now straightforward and results in 
\begin{eqnarray*}
{\cal A}_{R}(L,E;t) &=&\frac{d}{2}L^{d}\frac{4(\lambda -1)}{\pi ^{3/2}\sqrt{%
\lambda }\ln \lambda }\zeta \left\{ 1+O\left[ \exp (-\pi ^{2}/\zeta
^{2})\right] \right\} \\
&\simeq &\frac{d}{2}L^{d}\frac{8}{\sqrt{\pi }}\sqrt{t/t_{s}}\left\{
1+O\left[ \exp (-\pi ^{2}/\zeta ^{2})\right] \right\}
\end{eqnarray*}
where the {\em scale factor} for the short-time scaling is given by 
\begin{equation}
t_{s}(L,E)\equiv L^{d+2}\frac{4\sinh (LE/2)}{LE\exp (LE/2)}.  \label{te}
\end{equation}
While this characteristic scale obeys the same scaling form as the late
crossover time, (\ref{recall_lct}), 
\[
t_{s}(L,E)=L^{d+2}\tau _{s}(LE)\quad \text{with}\quad \tau _{s}(x)\equiv 
\frac{x\exp (x/2)}{4\sinh (x/2)} 
\]
we note that the scaling function $\tau _{s}(x)$ is different from $\tau
_{c}(x)$.


\newpage

\begin{references}
\bibitem{o+e}  This equation is strictly correct only for systems with an
odd number of sites. For a system with an even number of sites, the exact
relation is $N^{+}=N^{-}+1=\frac{N}{2}$, i.e., $N^{+}=N^{-}+O(L^{-2})$.
However, this correction will be neglected in the following: even for our
smallest system ($20\times 20$), it is below $1\%$.
\bibitem{new}  To the best of our knowledge, this is a new sum rule for
Bessel functions.
\end{references}
\end{document}